# Twenty-nine million Intrinsic Q-factor Monolithic Microresonators on Thin Film Lithium Niobate

XINRUI ZHU[1,†], YAOWEN HU[1,†], SHENGYUAN LU[1], HANA K. WARNER[1], XUDONG LI[1], YUNXIANG SONG[1], LETÍCIA MAGALHÃES[1], AMIRHASSAN SHAMS-ANSARI[1], NEIL SINCLAIR[1], AND MARKO LONČAR[1,*]

[1]*John A. Paulson School for Engineering and Applied Sciences, Harvard University, Cambridge, MA 02138, USA*
*†These authors contributed equally to this work*
*\*loncar@seas.harvard.edu*

**Abstract:** The recent emergence of thin-film lithium niobate (TFLN) has extended the landscape of integrated photonics. This has been enabled by the commercialization of TFLN wafers and advanced nanofabrication of TFLN such as high-quality dry etching. However, fabrication imperfections still limit the propagation loss to a few dB/m, restricting the impact of this platform. Here, we demonstrate TFLN microresonators with a record-high intrinsic quality (Q) factor of twenty-nine million, corresponding to an ultra-low propagation loss of 1.3 dB/m. We present spectral analysis and the statistical distribution of Q factors across different resonator geometries. Our work pushes the fabrication limits of TFLN photonics to achieve a Q factor within one order of magnitude of the material limit.

## 1. Introduction

Thin-film lithium niobate (TFLN) photonics is rapidly emerging as a versatile platform for high-speed electro-optics and nonlinear optics applications [1]. The material properties of lithium niobate (LN) lie at the center of this breakthrough: high refractive index, large $\chi^{(2)}$ nonlinearity, wide transparency window, and efficient piezo-electric response [2]. Recently, with commercialized TFLN wafers and advanced nanofabrication techniques based on high-quality dry etching [3], the exploration of lithium niobate photonics has been transformed from traditional bulk-scale components into integrated photonic chips, enabling both better performance and novel functionalities [1]. For example, integrated TFLN photonic devices, featuring strong nonlinear interaction, low propagation loss, and small form factors enable significant advances in electro-optic (EO) modulation [4, 5], EO frequency conversion [6], EO frequency comb generation [7, 8], Kerr frequency comb generation [9-12], synthetic crystal generation [13-15], transduction [16-19], and engineered all optical nonlinearities via periodic polling [20, 21]. These components represent fundamental building blocks for advanced large-scale photonic systems, boosting applications in optical communications [22], optical computation [23], precision measurements [24], microwave signal processing [25, 26], and quantum information science [27]. To unlock the full potential of the TFLN platform, propagation loss arises as a fundamental challenge. At the device level, enhancing light-matter interaction in various applications necessitates ultra-low propagation loss, as the ratio between coupling and loss determines the strength of the desired interaction. At the system level, the requirement of integrating numerous devices to form large-scale photonic circuits is ultimately limited by loss.

Substantial research efforts have been made to minimize the propagation loss of TFLN waveguides, that is maximize the quality (Q) factor of TFLN resonators. Using different resonator designs [28, 29], researchers have explored different etching methods to optimize fabrication quality, such as dry (ion-plasma) etching, wet etching, and chemical mechanical etching. Dry etching, utilizing reactive ion plasma, is the predominant method of etching LN

[3, 30-34] as it allows for precise control of feature sizes. Wet etching, though simpler, encounters the challenges of non-uniform etching along different crystal directions [35, 36]. While chemical mechanical polishing yields an ultra-smooth surface, it has difficulty in preserving narrow features and thus makes it hard to realize e.g. optical waveguide coupled resonators within a single TFLN layer [37-39]. Among works in monolithic TFLN, the highest recorded Q factor stands at 10 million, achieved through high-quality dry etching [3]. However, this achievement is notably below the optimal Q factors observed in other low-loss photonics platforms like 442 million on the silicon nitride platform [40] and the theoretical upper limit of 163 million inherent to the TFLN platform [41]. Furthermore, most breakthrough devices exhibit average Q factors limited to a few million [6-8], restricting the overall device performance. Consequently, the Q factor remains a critical bottleneck within the TFLN platform, requiring a further push on the upper limit.

Here, we demonstrate monolithic microring racetrack resonators on the TFLN platform with a record-high intrinsic Q factor of 29.3 million, and a corresponding ultra-low propagation loss of 1.3 dB/m, using the dry reactive ion etching. We achieved this by optimizing the etching process and tailoring the resonator widths and lengths to achieve the largest Q values. We evaluate the device fabrication quality using scanning electron microscopy (SEM) and atomic force microscopy (AFM). We also analyze the mode family from the spectrum of resonators, investigate the statistical relationship between the Q factor and resonator geometries, and calibrate the resonator linewidth using a radio-frequency (RF) modulated laser.

## 2. Device

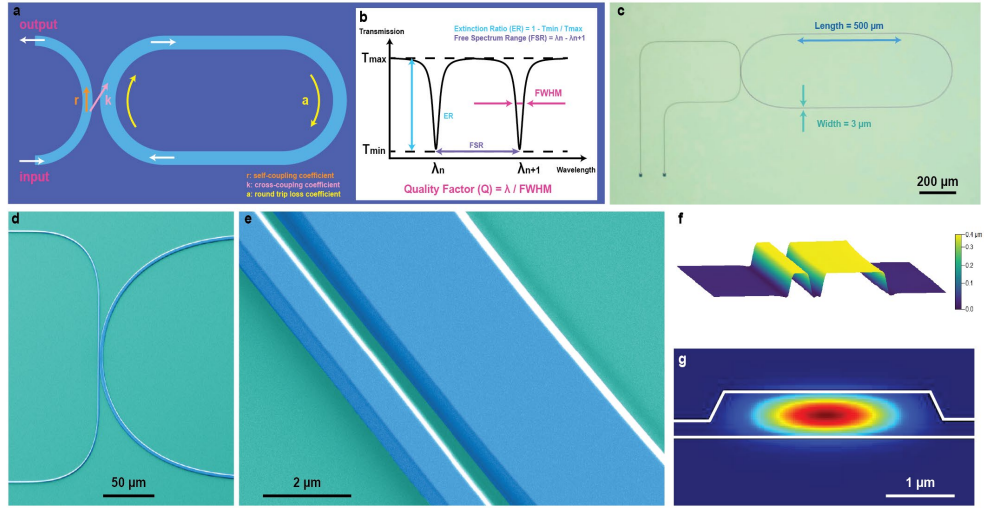

**Fig. 1. TFLN microresonator with smooth sidewall.** (a) Schematic illustration depicting the structure of a racetrack resonator and its light propagation mechanism. (b) Spectrum schematic demonstrating the characteristic features of a resonator. (c) Optical microscope image revealing a racetrack resonator with 3 μm width and 500 μm length racetrack. (d) SEM image offering an overview of the coupling region of a racetrack with 0.5 μm coupling gap and 3 μm width. (e) SEM image providing a detailed view of the coupling region of the same racetrack resonator. (f) AFM image capturing coupling region's topography. (g) Lumerical eigenmode simulation representing the fundamental TE mode at the cross-section of 3 μm width ring racetrack resonator.

Our devices employ a racetrack shape, forming a microresonator. As depicted in Figure 1a., the waveguide on the left can couple light in and out, and the cavity on the right is where light circulates for multiple roundtrips, experiencing the resonant enhancement. We design the racetrack width to be larger, aiming for a reduced overlap between the light and sidewall. Figure 1b. illustrates the anticipated spectrum of a ring resonator featuring a Lorentzian-shaped resonance at the specific resonant wavelength. The Q factor can be extracted from the linewidth and wavelength mathematically. A high Q factor implies a narrow resonance linewidth and minimal propagation loss.

We fabricate high Q resonators with an optimized fabrication process. The fabrication begins with a chip cleaved from a TFLN wafer with 600 nm thickness of LN and 4.7 μm buried silicon dioxide. The device is first patterned by electron beam lithography using Hydrogen Silsesquioxane (HSQ) as an electron beam resist. Subsequently, approximately 325 nm LN is etched using an inductively coupled plasma reactive ion etching tool with Argon gas. Following these steps, the device undergoes a chemical cleaning and annealing process. During the development of this fabrication process, we have tested individual steps, including, but not limited to, the type of resist, exposure dose, and etching conditions (plasma power, etching time, etching depth). This optimized fabrication method exhibits stability and repeatability in consistently yielding functional devices.

We have conducted characterizations on our racetrack resonator devices to verify the fabrication quality. Figure 1c. presents the optical microscope image of a racetrack resonator with length of 500 μm and bending radius of 200 μm. Figure 1d-e. are the scanning electron microscopy (SEM) images of a coupling region of 3 μm racetrack width device with a coupling gap of 0.5 μm. Taken by atomic force microscopy (AFM) image, Figure 1f. is consistent with Figure 1e, demonstrating the well-defined waveguide and smooth sidewalls, suggesting high fabrication quality. Additionally, Figure 1g, obtained from Lumerical simulation, illustrates the fundamental mode supported by a 3 μm width racetrack, showing the reduced overlap between the light and sidewall enabled by the ultra-wide waveguide.

## 3. Measurement

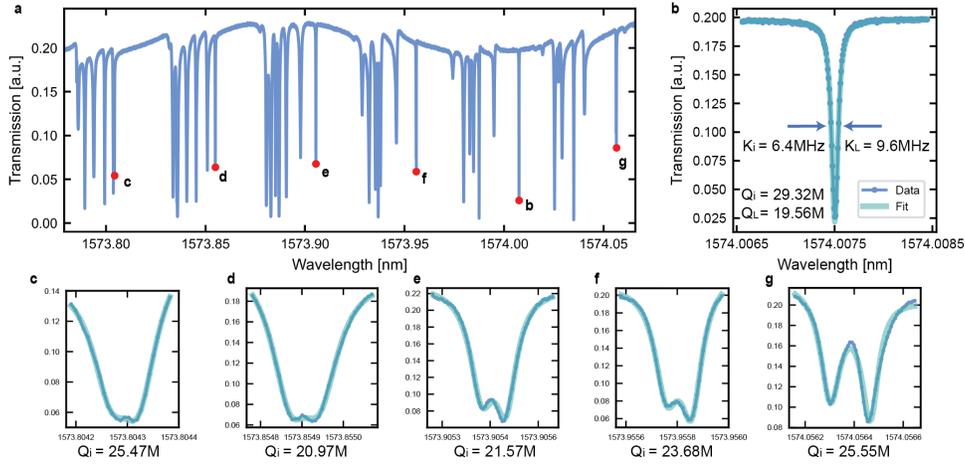

**Fig. 2. Monolithic high Q microresonators on TFLN.** (a) Selected resonator spectrum spanning from wavelength 1573.78 nm to 1574.06 nm. The corresponding racetrack features a width of 4.5 μm, length of 10 mm, coupling gap of 0.6 μm, and bending radius of 200 μm. Background modulation is attributed to the cavity formed by reflections between the two facets of the chip. (b) The highest Q resonance features an intrinsic Q factor of 29 million at the wavelength of 1574 nm. (c) - (g) Resonances at wavelength 1573.8 nm, 1573.85 nm, 1573.90 nm, 1573.95 nm, and 1574.04 nm, all belonging to the same high Q mode family.

We measure our device using a tunable laser and a low-noise detector. Figure 2a. demonstrates an exemplary spectrum obtained from our measurements, where the x-axis represents the wavelengths generated through laser sweep, and the y-axis denotes the signal received by the detector. The spectrum presents diverse resonance shapes with noticeable linewidth and extinction ratio distinctions because the wide racetracks support both fundamental and higher-order modes. Different mode families are distinguished by picking up modes with similar shapes and extinction ratios and fitting their free spectral range (FSR). Fundamental modes are selected by finding the mode family with the highest FSR (depicted by the red dots in Figure 2a.), corresponding to the minimum group index predicted by the Lumerical eigenmode simulation. We observe that the fundamental mode exhibits the highest Q factor among all the mode families (referred to as "high Q mode family" later), which is consistent with our design as the fundamental mode typically has the smallest amount of overlap with waveguide sidewall roughness.

Many measured resonances feature ultra-high Q factors. The highest intrinsic Q factor is 29.32 million, with corresponding loaded Q factor of 19.56 million, corresponding to a device with parameters of 4.5 μm width, 10 mm length, and 0.6 μm coupling gap (Figure 2b). The intrinsic and loaded linewidth are 6.4 MHz and 9.6 MHz, respectively. Other resonances within the same mode family exhibit ultra-high Q values exceeding 20 million, as shown in Figure 2c-d. Resonances from the high Q mode family may feature single or split peaks, which could have resulted from the mode splitting between the clockwise (CW) and counter-clockwise (CCW) modes. The typical CW and CCW mode coupling are small; however, in this ultrahigh Q regime, the linewidth could be comparable to this coupling, leading to the observation of mode splitting. In addition, the potential overlap with resonances from other mode families may also generate similar doublet features.

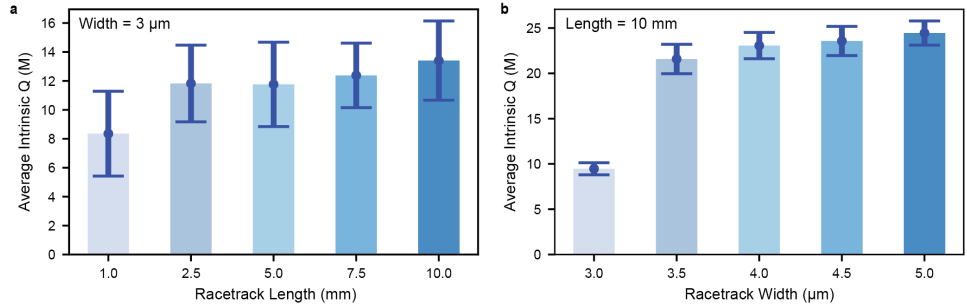

Fig. 3. **Statistical analysis of intrinsic Q factor versus racetrack length and width.** The bars depict the mean value of the top 50 intrinsic Q-factors of a device, while horizontal marker lines are error bars, indicating the mean values plus and minus their standard deviations. (a) Average intrinsic Qs of racetracks with width of 3 μm and length of 1 mm, 2.5 mm, 5 mm, 7.5 mm, and 10 mm. (b) Average intrinsic Qs of racetracks with a length of 10 mm and width of 3.0 μm, 3.5 μm, 4.0 μm, 4.5 μm, and 5.0 μm.

We perform a comprehensive statistical analysis of Q factors by examining the entire scan range of wavelengths (1480 nm to 1620 nm). For each device, an automated Python code searches for all the resonances and fits their Q values. We select the top 50 resonances from each device to represent the high Q mode family and evaluate their average value and standard deviation. As illustrated in Figure 3, we compare the statistical Q result for devices with different geometry. For the devices with 3 μm width, we vary the length from 1 mm to 10 mm, revealing an increasing Q trend. In the devices with 10 mm length on another chip, the Q increases as waveguide width increases from 3 μm to 5 μm. While we extract our highest Q record from a racetrack with 4.5 μm width and 10 mm length, the average Q values of the device with 5 μm width and 10 mm length is slightly better. Furthermore, the two chips for Figure 3a. and Figure 3b. are made in two individual fabrication rounds, affirming the reproducibility of our high Q results.

## 4. Calibration

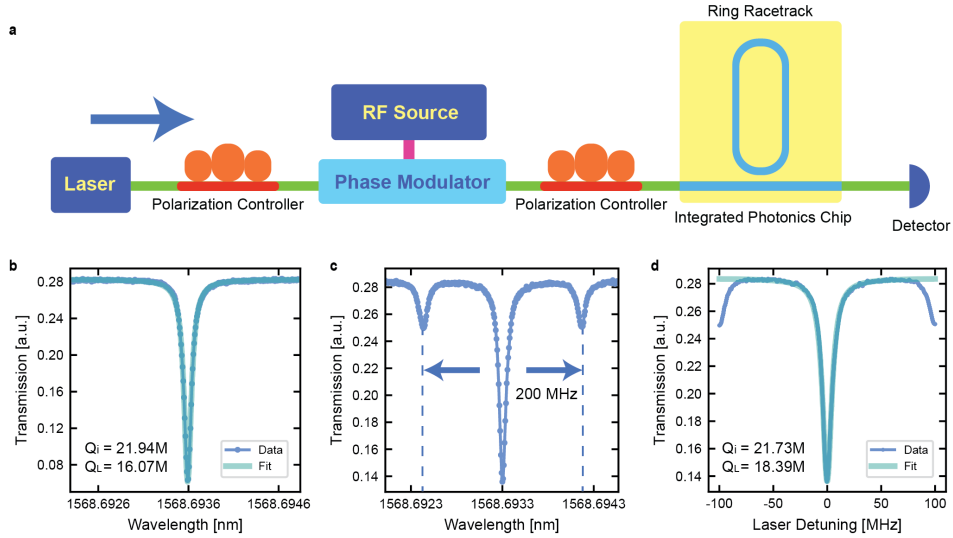

**Fig. 4. Resonance calibration with RF-modulated laser light.** (a) Schematic of measurement setup incorporating a phase modulator capable of generating optical sidebands. (b) The original resonance without applying RF power exhibits a loaded Q of 16.07 million. (c) Same resonance with sidebands generated by activating the 100 MHz RF sources. (d) Calibration using the sidebands positions to redefine the x-axis as frequency and refit the resonance. The loaded Q increases to 18.39 million, slightly higher than the original loaded Q.

To verify that the direct laser sweep approach can characterize the actual spectrum precisely, we further evaluate the measurement using RF modulation as a calibration. Figure 4a. depicts a phase modulator driven by an RF source that is added before the resonator to generate sidebands. The distance between sideband and the original signal is used as a ruler to calibrate the x axis. Figure 4b-d. demonstrate the calibration process. Initially, we measure a resonance without RF modulation. Immediately after, we turn on the RF modulation with an RF frequency of 100 MHz and observe the generation of sidebands. With sidebands as a reference, we calibrate the x-axis, extract the linewidth, and calculate Q values. It is worth noting that although this approach allows one to calibrate the linewidth using a precise RF frequency reference, it underestimates the extinction ratio: sidebands generated by RF signal have non-zero transmission when the laser signal is on resonance with the cavity. The fitting, using simple Lorentzian, then underestimates effective "coupling loss" and thus underestimating intrinsic Q. Care needs to be taken to compensate for this effect. Taking this into account, we evaluate that perceived RF-calibrated loaded Q of 18.39 million, which is comparable to the original loaded Q of 16.07 million.

## 5. Conclusion

We demonstrate microresonators with a record-high intrinsic quality factor of 29 million on monolithic TFLN. We performed statistical analysis of measured quality factor over different device designs and calibrated the measurement with an RF modulated laser. The optimized design and fabrication process of monolithic TFLN microresonators enable our realization of ultrahigh Q devices. Moving forward, the enhanced high Q devices on TFLN could substantially improve the device performances of electro-optics and nonlinear photonics, transitioning devices into a new parameter space. This advancement holds great potential to catalyze system-level applications, thus facilitating applications in microwave photonics [42], quantum computing [43, 44], and nonlinear optics [45, 46]. Our efforts further push the state-of-the-art, showcasing the potential of the TFLN platform and paving the way for future innovative explorations in integrated photonics.

**Funding.** Defense Advanced Research Projects Agency (HR001120C0137); U.S. Navy (N68335-22-C-0413); Air Force Office of Scientific Research (FA9550-20-1-01015); Air Force Research Laboratory (FA864921P0781); National Aeronautics and Space Administration (80NSSC22K0262, 80NSSC23PB442); National Science Foundation (EEC-1941583, OMA-2137723, 2138068); Office of Naval Research (N00014-22-C-1041); National Institutes of Health (5R21EY031895-02); National Research Foundation of Korea; S.L. acknowledges fellowship from Agency for Science, Technology and Research (A-STAR).

**Disclosures.** ML: HyperLight Corporation (F,I,C).